\documentstyle[prl,aps,psfig]{revtex}

\newcommand{\be}{\begin{equation}}
\newcommand{\ee}{\end{equation}}
\newcommand{\fig}[1]{fig.~\ref{#1}}

\begin{document}

\draft

\twocolumn[\hsize\textwidth\columnwidth\hsize\csname@twocolumnfalse\endcsname

\title{Persistence of Kardar-Parisi-Zhang Interfaces}
\author{Harald Kallabis$^{(1,2)}$ and Joachim Krug$^{(3)}$}

\address{
(1) Center for Polymer
Studies, Boston University, Boston, MA 02215, USA \\ 
(2) H\"ochstleistungsrechenzentrum, Forschungszentrum	
J\"ulich, 52425 J\"ulich, Germany\\
(3) 
Fachbereich Physik, Universit\"at GH Essen, D-45117 Essen, Germany
}

\maketitle

\begin{center}
\today
\end{center}

\begin{abstract}
The probabilities $P_\pm(t_0,t)$ that a growing Kardar-Parisi-Zhang
interface remains above
or below the mean height in the time interval $(t_0, t)$ are shown
numerically to decay as $P_\pm \sim (t_0/t)^{\theta_\pm}$ with
$\theta_+ = 1.18 \pm 0.08$ and $\theta_- = 1.64 \pm 0.08$.  Bounds on
$\theta_\pm$ are derived from the height autocorrelation function
under the assumption of Gaussian statistics.  The autocorrelation
exponent $\bar \lambda$ for a $d$--dimensional interface with
roughness and dynamic exponents $\beta$ and $z$ is conjectured to be
$\bar \lambda = \beta + d/z$.  For a recently proposed discretization
of the KPZ equation we find oscillatory persistence probabilities,
indicating hidden temporal correlations.
\end{abstract}

]

\narrowtext

The notion of persistence plays a central role in the assessment of
random processes in science and everyday life. Intuitively, a process
is persistent if it tends to maintain its current trend. For example,
if a continuous random process with zero mean is found at a positive
value at time $t_0$, one may ask for the {\em persistence probability}
$P_+(t_0,t)$ that it remains above zero up to time $t > t_0$; the
probability $P_-(t_0,t)$ for remaining negative is defined
analogously.

Here we consider the persistence of growing interfaces.  The random
variable of interest is the height $h(x,t)$ above some fixed but
arbitrary substrate site $x$, relative to the mean height $ \bar
h(t)$. At time $t=0$ the interface is flat, $h(x,t) = 0$, and we wish
to characterize $P_\pm(t_0,t)$ for $0 < t_0 < t$; in general positive
and negative persistence probabilities differ, because the growth
breaks the $h \to -h$ symmetry.  The dynamic scale invariance of
interface growth \cite{reviews} implies a scaling form for the
temporal autocorrelation function,

\begin{equation}
\label{auto}
A(t,t') \equiv \langle (h(x,t) - \bar h(t))(h(x,t') - \bar h(t')) \rangle
= t^{2\beta} {\cal A}(t'/t),
\end{equation}
where $\beta > 0$ is the temporal roughness exponent; in particular,
the surface width $W(t) = \sqrt{A(t,t)}$ grows as $t^\beta$. By the
same token, the persistence probabilities $P_\pm(t_0,t)$ can depend
only on the ratio of the two time arguments, and for $t \gg t_0$ one
expects a power law decay

\begin{equation}
\label{thetas}
P_{\pm}(t_0,t) \sim (t_0/t)^{\theta_\pm},
\end{equation}
defining a pair of persistence exponents $\theta_\pm$. 

A natural question concerns the relationship, if any, between the
roughness exponent $\beta$ and the persistence exponents
$\theta_{\pm}$. If $\beta$ is large, the height quickly deviates from
the mean and is unlikely to return; thus, $\theta_\pm$ should be
small. Indeed, when the initial condition $h(x,0)$ is an interface
with fully developed (stationary) roughness, the translational
invariance in time can be exploited to derive the simple relationship
\cite{five,rio}

\begin{equation}
\label{stationary}
\theta_+^{(s)} = \theta_-^{(s)} \equiv \theta^{(s)} = 1 - \beta,
\end{equation}
where the superscript refers to persistence in the steady state.  For
the flat initial condition a detailed study \cite{five} of {\em
linear} growth equations (for which $\theta_+ = \theta_- \equiv
\theta_0$) shows that $\theta_0$ still increases with decreasing
$\beta$, however the precise dependence is nontrivial and only
accessible to perturbation theory and rigorous bounds. In fact the two
exponents $\theta^{(s)}$ and $\theta_0$ describe different asymptotic
regimes of a single generalized persistence probability \cite{five}.

In the present paper we extend the work of Ref.\cite{five} to the
generic nonlinear interface growth equation \cite{reviews}

\begin{equation}
\label{kpz}
\frac{\partial h}{\partial t} = \nu \nabla^2 h + 
\frac{\lambda}{2} (\nabla h)^2 + \eta(x,t)
\end{equation}
introduced by Kardar, Parisi and Zhang (KPZ) \cite{kpz}.  This entails
several complications. First, we expect (and verify numerically)
different values for $\theta_+$ and $\theta_-$. Second, due to the
nonlinearity of (\ref{kpz}), the height fluctuations are non-Gaussian
and therefore the autocorrelation function (\ref{auto}), some
properties of which are known \cite{jk91,kmhh,krech,diss}, does not
even in principle fully determine the persistence probability. The
techniques developed in \cite{five} are used to derive bounds on the
persistence exponent of a Gaussian process with the KPZ
autocorrelation function. Our numerical estimates violate the bounds,
indicating that the non-Gaussian nature of the interface fluctuations
is important. Apart from its intrinsic interest, the investigation of
persistence is further motivated by the observation, to be reported
below, that $P(t_0,t)$ is a sensitive probe for hidden temporal
correlations in the growth process. This could shed some light on
recent claims of nonuniversal scaling in discrete models associated
with (\ref{kpz}) \cite{tim}.

{\em Numerical results.}  Simulations were carried out in one
dimension, using the strong-coupling discretization of the KPZ
equation proposed by Newman and Swift (NS) \cite{tim} as well as the
restricted solid-on-solid (RSOS) growth model of Kim and Kosterlitz
\cite{kk}.  The NS model is a discrete-time model with continuous
height variables $h(x,t)$ evolving in parallel according to

\begin{equation}
\label{ns}
h(x,t+1) = \max[h(x,t),h(x-1,t),h(x+1,t)] + \eta(x,t),
\end{equation}
where the noise variables $\eta(x,t)$ are drawn from a 
Gaussian distribution with unit variance; the effect of using
other distributions will be discussed at the end of the paper.

\begin{figure}[htb]
\centerline{\psfig{figure=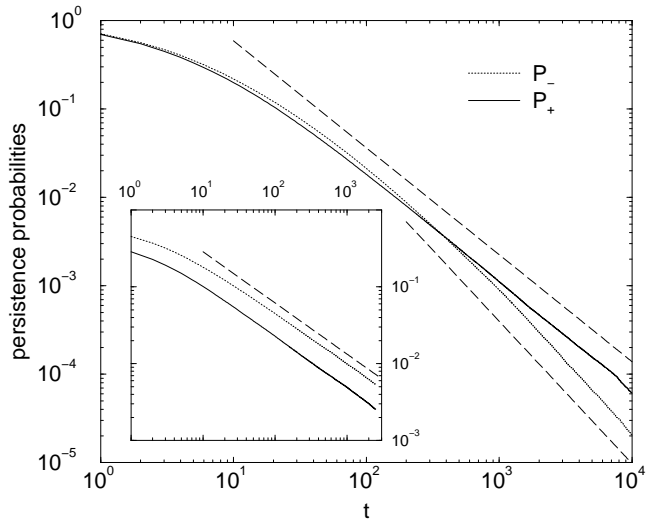,height=7cm,angle=270}}
\caption{
Persistence probabilities for the NS model with Gaussian noise.  The
dashed lines have the slopes $-1.21$, and $-1.61$, respectively.
Inset: Steady state persistence probabilities obtained for a rough
surface as initial condition ($t_0=10^4$ time steps, starting from a
flat surface at time zero).  The dashed line has the slope $-2/3$.
}
\label{NSpers.fig}
\end{figure}

We first checked relation (\ref{stationary}) for growth from a rough
initial condition. Using the NS model we find $\theta_+^{(s)} =
\theta_-^{(s)} = 0.66 \pm 0.03$ (see the inset of \fig{NSpers.fig}),
which is in good agreement with (\ref{stationary}) and the exact value
$\beta = 1/3$ of the one-dimensional KPZ equation \cite{kpz}. This
confirms the conjecture \cite{five,rio} that the validity of
(\ref{stationary}) is not restricted to Gaussian processes.

To determine the nontrivial exponents $\theta_{\pm}$, the interface
was started in the flat state, $h=0$, and the persistence was measured
from $t_0 = 1$, i.e.\ one time step in the NS model and one deposition
attempt per site in the RSOS model.  The system size in the NS model
was $L=16384$ and averages were taken over 1000 independent runs. The
measurements shown in \fig{NSpers.fig} yield the estimates $\theta_+ =
1.21 \pm 0.06$, $\theta_-=1.61 \pm 0.08$ which were obtained from fits
in the regions $100\le t \le 6000$ and $600\le t\le 6000$,
respectively. The error bars are based on the fluctuations of the
running exponent.  The fact that $\theta_+ < \theta_-$ is made
plausible by noting that the nonlinear term $(\lambda/2)(\nabla h)^2 $
in (\ref{kpz}) describes lateral growth, i.e.\ it causes, for $\lambda
> 0$, humps on the surface to extend sideways \cite{reviews,kpz}. This
tends to push negative segments of the surface (where $h < \bar h$)
upwards and leads to a faster decay of $P_-$.  Since $\lambda>0$ in
the NS model \cite{unpub}, this argument explains the inequality. It
also suggests a faster decay of the height probability distribution in
the direction of positive heights when $\lambda < 0$, in agreement
with numerical \cite{kim91} and exact \cite{derrida98} results on
related models.

For the RSOS model the persistence probability was averaged over 124
runs, using a system size $L = 5 \times 10^5$ and $10^4$ growth
attempts per site. A fit to the data for times $500 \leq t \leq 5000$
yields $\theta_+ = 1.67 \pm 0.07$ and $\theta_- = 1.15 \pm 0.08$, with
errors estimated from the observed fluctuations in the running
exponent; the RSOS model has $\lambda < 0$ \cite{kscomm}, hence
$\theta_+ > \theta_-$.  Within error bars the exponents agree with
those of the NS model.  The numbers reported in the abstract represent
the average of the values obtained for the two models.

\begin{figure}[htb]
\centerline{\psfig{figure=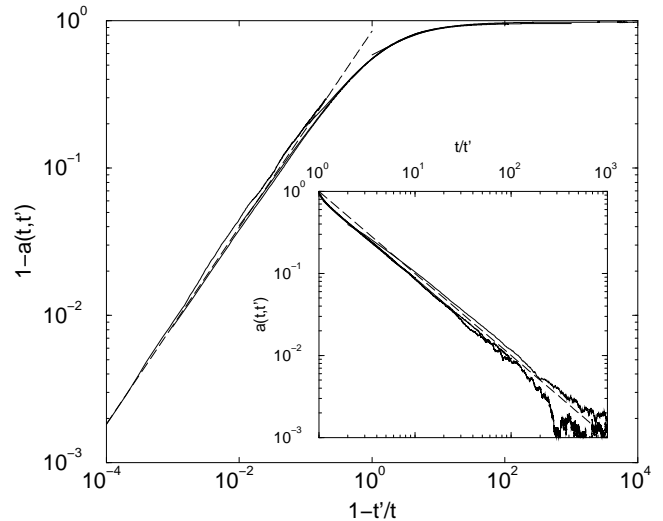,height=7cm,angle=270}}
\caption{
The early-time behavior $1-a(t,t')=(R/2)(1-t'/t)^{2/3}$ of the
autocorrelation functions for $t'=10^0,10^1,10^2,10^3,10^4$.
The dashed line is the curve $0.85\cdot(1-t'/t)^{2/3}$. Inset:
The late-time behavior $a(t,t')\propto (t/t')^{-1}$ of the
autocorrelation function for $t'=10^0,10^1,10^2,10^3,10^4$.  The
dashed line is $(t/t')^{-1}$.
}
\label{a1.fig}
\end{figure}

{\em Autocorrelation function.} For a quantitative understanding of the
persistence exponents we need to characterize the shape of the KPZ
autocorrelation function (\ref{auto}). It is useful to introduce two
related functions: The temporal height difference correlation function

\begin{equation}
\label{C}
C(t,t') = \langle (h(x,t) - h(x,t') - \bar h(t) + \bar h(t') \rangle^2 
\end{equation}
and the normalized autocorrelation function 

\begin{equation}
\label{a}
a(t,t') = \frac{A(t,t')}{W(t) W(t')} = (t/t')^\beta {\cal A}(t'/t).
\end{equation}
The height difference function can be shown to have the form
\cite{kmhh,diss}

$
C(t,t') = c(t'/t) \vert t - t' \vert^{2\beta}
$
where the amplitude $c(t'/t)$ tends to constants $c_2$ and $\tilde
c_2$ for $t'/t \to 0$ and $t'/t \to 1$, respectively (we take $t >
t'$).  Their ratio $R = \tilde c_2 / c_2$ is a universal number, whose
value is numerically estimated \cite{kmhh,diss} to be $R = 1.8 \pm
0.1$. Inserting this into (\ref{a}) it follows that $a(t,t')
\approx 1 - (R/2)(1 - t'/t)^{2/3}$ for $t'/t \to 1$ (\fig{a1.fig}).

For $t \gg t'$ the normalized autocorrelation function decays as a
power law, $a(t,t') \sim (t'/t)^{\bar \lambda}$, where the
autocorrelation exponent $\bar \lambda = 1$ \cite{jk91,krech,diss}
(\fig{a1.fig}). This is a special case of the general identity

\begin{equation}
\label{lambda}
\bar \lambda = \beta + d/z
\end{equation}
relating $\bar \lambda$ to the surface dimensionality $d$, the
roughness exponent $\beta$ and the dynamic exponent $z$; for the
one-dimensional KPZ equation $z = 3/2$ \cite{kpz}. Eq.(\ref{lambda})
is valid for the linear growth equations considered in \cite{five},
and we conjecture that it holds generally for rough growing
interfaces.  To see this, note that $a(t,t')$ measures the overlap
between the height configurations at times $t$ and $t'$. The overlap
is the product of two factors: The lateral overlap between domains in
the $d$-dimensional substrate space, which is of the order
$(\xi(t')/\xi(t))^d$ with the dynamic correlation length
\cite{reviews} $\xi(t) \sim t^{1/z}$, and the horizontal overlap
$W(t')/W(t) \sim (t'/t)^\beta$.

{\em Analytic bounds.}
Having established the behavior of the autocorrelation function for
short and long times, the method of Ref.~\cite{five} can be used to
obtain an estimate for the persistence exponents
$\theta_{\rm KPZ}$ of the Gaussian stochastic
process associated with the KPZ autocorrelation function 
$a_{\rm KPZ}(t,t')$.
We first make the process stationary by passing to
logarithmic time \cite{perspapers}. Setting $T = \ln t$ we obtain
$a_{\rm KPZ}(t,t') = f_{\rm KPZ}(T - T')$, where the function $f_{\rm
KPZ}$ has the limiting behaviors

\begin{equation}
\label{flimit}
f_{\rm KPZ}(T) \approx \left\{ \begin{array}{l@{\quad:\quad}l}
 1 - (R/2) \vert T \vert^{2/3}  &  T \to 0 \\ 
 e^{-T} &  T \to \infty, \end{array} \right.
\end{equation}
and the persistence probability decays exponentially, as
$e^{-\theta_{\rm KPZ} T}$. The function $f_{\rm KPZ}$ will be compared
to the logarithmic time autocorrelation function $f_\beta(T)$ of the
{\em linear} growth equation with roughness exponent $\beta = 1/3$,
which is given by \cite{five}

\begin{equation}
\label{linear}
f_\beta(T) = \cosh(T/2)^{2\beta} - \vert \sinh(T/2) \vert^{2\beta}.
\end{equation}
The value $\beta = 1/3$ can be realized e.g.\ by the linearized
version of (\ref{kpz}) with spatially correlated noise. For short and
long times $f_\beta$ behaves as $1 - 2^{-2\beta} \vert T
\vert^{2\beta}$ and $e^{-(1 - \beta)T}$, respectively. Thus, at least
in these limiting regimes, we have $f_{\rm KPZ} < f_{1/3}$. If
this is true for all $T$, a comparison theorem due to Slepian
\cite{slepian} states that the persistence probabilities of the
Gaussian processes associated with $f_{\rm KPZ}$ and $f_{1/3}$ satisfy
the same inequality, and hence $\theta_{\rm KPZ} >
\theta_{1/3}$. In \cite{five} it was shown that $\theta_{1/3} \geq
(2/3) 2^{1/2} \approx 0.942809...$, while the numerical solution
of the linear growth equation yields $\theta_{1/3} = 0.95 \pm 0.05$. 

More precise bounds can be derived by determining optimal scaling
parameters $b_{\rm max}$ and $b_{\rm min}$ such that the inequalities
$f_{1/3}(b_{\rm max} T ) \leq f_{\rm KPZ}(T) \leq f_{1/3}(b_{\rm min}
T ) $ hold for all $T$ \cite{five}; Slepian's theorem then implies
$b_{\rm min} \theta_{1/3}\leq \theta_{\rm KPZ}\leq b_{\rm
max}\theta_{1/3}$. Comparing (\ref{flimit}) and (\ref{linear}) and
using the numerical estimate $R \approx 1.8$ we obtain the necessary
conditions 
$ b_{\rm max} \geq 2 (R/2)^{3/2} \approx 1.7$ and 
$ b_{\rm min} \leq 3/2$. 
Inspection of the numerically determined function $a_{\rm KPZ}$
yields the optimal parameters 
$b_{\rm min} = 1.35$ and $b_{\rm max} = 4$,
hence we conclude 
$1.28 \leq \theta_{\rm KPZ} \leq 3.8$. 
The fact that our numerical estimate for $\theta_+$ (which dominates
the {\em total} persistence probability $P_+ + P_-$ for long times) weakly
violates the lower of these bounds appears to be a manifestation
of the non-Gaussian character of the KPZ interface fluctuations.

\begin{figure}[htb]
\centerline{\psfig{figure=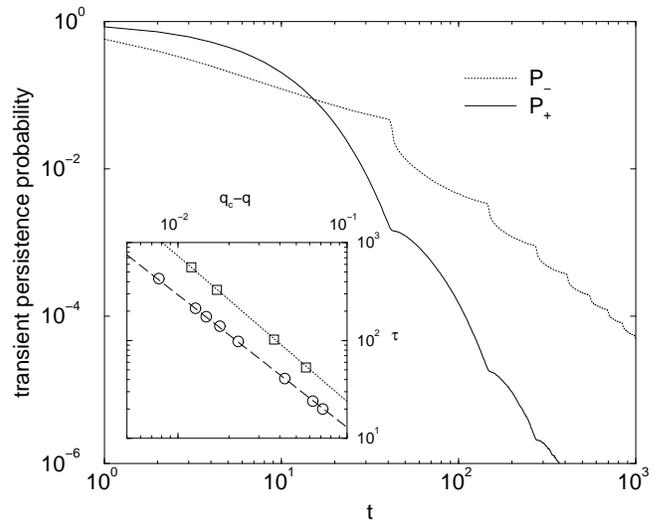,height=7cm,angle=270}}
\caption{
Persistence probabilities in the NS model with symmetric
discrete noise. Inset: Period of the first (circles) and
second (squares) oscillation as a function of $q_c-q$.  Using
three-parameter fits we obtain $q_c=0.542$ in the first and
$q_c=0.537$ in the second case.  The dashed and the dotted line have
the slopes $-1.355$ and $-1.488$, respectively.
}
\label{NSbin.fig}
\end{figure}

{\em Non-Gaussian noise.}
Newman and Swift \cite{tim} recently reported a remarkable dependence
of the roughness exponent of high-dimensional KPZ growth on the
probability distribution $p(\eta)$ of the noise variable $\eta$ in
(\ref{ns}).  Since the persistence probability (in contrast e.g.\ to
the surface width or the autocorrelation function) probes temporal
correlations of arbitrary order \cite{slepian}, it seemed natural to
ask whether similar effects could be identified in low dimensions
using $P_\pm(t_0,t)$ as a dynamic observable.

In \fig{NSbin.fig} we show measurements of the persistence
probabilities for the NS model with a symmetric, discrete noise
distribution $p(\eta) = (1/2)[\delta(\eta - 1) + \delta(\eta + 1)]$.
Instead of the expected power law decay (\ref{thetas}), the $P_{\pm}$
show distinct oscillations, indicative of an intrinsic time scale
$\tau \approx 100$; we emphasize that no oscillations were found in
the surface width or other more conventional quantities. The origin of
the time scale $\tau$ can be explained by considering the general
class of distributions
$
p_q(\eta) = q\delta(\eta - 1) +  (1-q)\delta(\eta + 1)
$
where $0 < q < 1$. With this noise distribution the NS model (\ref{ns})
becomes equivalent to the polynuclear growth (PNG) model
studied by Kert\'esz and Wolf \cite{kw}, which is known to undergo a
directed percolation transition at $q_c \approx 0.539$; for $q > q_c$
the surface does not roughen. The transition is associated with a
diverging correlation time $\tau \sim \vert q - q_c \vert^{-\nu_t}$,
where $\nu_t \approx 1.73$ is the temporal correlation length exponent
of directed percolation. Measurements of the period of oscillations in
$P_\pm$ as a function of $q$ show a similar power law divergence,
confirming the identification of $\tau$ as the directed
percolation correlation time (inset of \fig{NSbin.fig}). 

We have found similar, though less pronounced oscillations for the
continuous, bimodal noise distributions \cite{tim}
$ p_\alpha(\eta) \sim (1-|\eta|)^\alpha $
with $\alpha > -1$. Further work is required
to clarify the role of temporal correlations (induced by the parallel
updating scheme (\ref{ns})) in giving rise to the non-universal 
exponents observed by NS. The problem may indeed become more severe
in higher dimensions due to the decrease of the directed
percolation threshold $q_c$; in 2+1 dimensions $q_c \approx 0.272$
\cite{kw}, hence for the symmetric discrete distribution ($q = 1/2$)
the NS model resides in the smooth phase.   

{\em Acknowledgements.}
J.K. acknowledges useful discussions with Alan Bray and financial
support by DFG within SFB 237 H.K. acknowledges support by the German
Academic Exchange Service within the Hochschulsonderprogramm III.

\end{document}